\documentclass{article}

\usepackage{microtype}
\usepackage{graphicx}
\usepackage{subcaption}
\usepackage{booktabs} %

\usepackage{hyperref}

\usepackage[preprint]{icml2026}

\usepackage{amsmath}
\usepackage{amssymb}
\usepackage{mathtools}
\usepackage{amsthm}

\usepackage[capitalize,noabbrev]{cleveref}

\theoremstyle{plain}

\theoremstyle{definition}

\theoremstyle{remark}

\usepackage[disable,textsize=tiny]{todonotes}

\usepackage{epigraph}
\usepackage{enumitem}
\usepackage{stfloats}
\usepackage{comment}
\usepackage{wrapfig}
\usepackage{listings}

\definecolor{ColorGray}{RGB}{180, 180, 180}
\definecolor{ColorDarkGreen}{RGB}{19, 179, 50}
\definecolor{ColorLightBlue}{RGB}{18, 137, 255}
\definecolor{ColorOrange}{RGB}{252, 111, 3}

\usepackage{pifont}

\def\eg{\emph{e.g.}}

\def\ie{\emph{i.e.}}

\def\wrt{w.r.t.}

\newcommand{\mypara}[1]{\noindent\textbf{#1}~}

\def\OURS{HOI-PAGE}

\def\TextPrompt{\varGamma}

\def\Object{\mathcal{O}}
\def\ObjectRotation{\mathbf{R}}
\def\ObjectTranslation{\mathbf{t}}

\def\Human{\mathcal{H}}
\def\HumanParameters{\Theta}

\def\MotionLength{T}

\def\Graph{\mathcal{G}}
\def\GraphNode{\mathbf{v}}
\def\GraphNodes{\mathcal{V}}
\def\GraphVirtualNode{\overline{\mathbf{v}}}
\def\GraphNodeRotational{a_{r}}
\def\GraphNodeTranslational{a_{\tau}}
\def\GraphEdge{\mathbf{e}}
\def\GraphEdges{\mathcal{E}}
\def\GraphEdgeContinuous{a_{c}}
\def\GraphEdgeStatic{a_{s}}
\def\GraphEdgeNodePointCloud{\mathcal{P}}

\def\VideoFrame{I}

\def\LossFit{\mathcal{L}_{\text{fit}}}
\def\LossFitWeight{\lambda_{\text{fit}}}
\def\LossObjectFit{\mathcal{L}^{\Object}}
\def\LossObjectPartFit{\mathcal{L}^{\text{o}}}
\def\LossContact{\mathcal{L}_{\text{con}}}
\def\LossContactWeight{\lambda_{\text{con}}}
\def\LossContactContinuity{\mathcal{L}_{\text{cc}}}
\def\LossContactDynamics{\mathcal{L}_{\text{cd}}}
\def\LossPenetration{\mathcal{L}_{\text{pen}}}
\def\LossPenetrationWeight{\lambda_{\text{pen}}}
\def\LossSmoothness{\mathcal{L}_{\text{smo}}}
\def\LossSmoothnessWeight{\lambda_{\text{smo}}}
\def\LossSmoothnessRotation{\mathcal{L}_{\text{r}}}
\def\LossSmoothnessTranslation{\mathcal{L}_{\tau}}
\def\LossTotal{\mathcal{L}_{\text{total}}}

\def\LossEuclidean{\mathcal{L}_{2}}

\icmltitlerunning{\OURS{}}

\begin{document}

\twocolumn[
  \icmltitle{\OURS{}: Zero-Shot Human-Object Interaction Generation with Part Affordance Guidance}

  \icmlsetsymbol{equal}{*}
  \icmlsetsymbol{tumprev}{$\dagger$}

  \begin{icmlauthorlist}
    \icmlauthor{Lei Li}{uva,tumprev}
    \icmlauthor{Angela Dai}{tum}
  \end{icmlauthorlist}

  \icmlaffiliation{uva}{University of Virginia, Charlottesville, VA, United States}
  \icmlaffiliation{tum}{Technical University of Munich, Munich, Germany}

  \icmlcorrespondingauthor{Lei Li}{leili@virginia.edu}

  \icmlkeywords{4D Human-Object Interaction Synthesis, Part Affordance from Large Language Models, Zero-Shot HOI with Video Diffusion Priors}

  \vskip 0.3in

{
\begin{center}
    \centering
    \captionsetup{type=figure}
    \includegraphics[width=\textwidth]
    {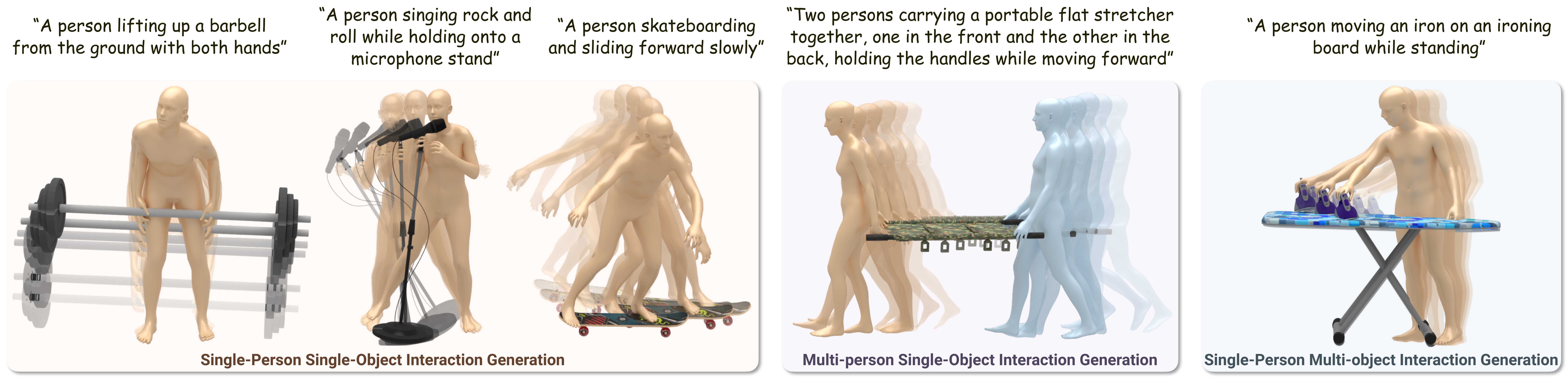}
    \captionof{figure}{We propose to model complex 4D human-object interactions (HOIs), including those involving multiple objects or people, by inferring part affordance graphs (PAGs) that guide zero-shot HOI synthesis from a text prompt and 3D object model(s).
Our PAGs, distilled from large language model reasoning, provide localized affordance constraints for our optimization-based generation, enabling flexible modeling of diverse interaction scenarios in a zero-shot fashion.}
    \label{fig_teaser}
\end{center}%
\vskip 0.3in
}

]

\printAffiliationsAndNotice{\textsuperscript{$\dagger$}Work partly done while Lei Li was a postdoc at TUM.}  %

\begin{abstract}
  We present \OURS{}, a new approach that prioritizes part-level affordance reasoning to generate high-fidelity 4D human-object interactions (HOIs) from text prompts in a zero-shot fashion. In contrast to prior works that focus on global, whole body-object motion synthesis, our approach explicitly reasons about the underlying part-level mechanics of interactions using large language models (LLMs). We capture this reasoning in a structured part affordance graph (PAG) representation, serving as a high-level interaction scaffolding to guide a three-stage synthesis: first, decomposing input 3D objects into semantic parts; then, generating reference HOI videos from text prompts to extract part-based motion constraints; and finally, optimizing for 4D HOI motion sequences that mimic the reference dynamics while satisfying part-level contact constraints. Extensive experiments show that our approach is flexible and capable of generating complex multi-object or multi-person interaction sequences, with significantly improved realism and text alignment for zero-shot 4D HOI generation.
\end{abstract}

\section{Introduction}
\label{sec_intro}

\epigraph{
``The \textit{affordances} of the environment are what it \textit{offers} the animal, what it \textit{provides} or \textit{furnishes}, either for good or ill. ... It implies the complementarity of the animal and the environment.''
}{-- James J. Gibson}

Human-object interaction (HOI) is a fundamental aspect of everyday life, ranging from simple activities like picking up a cup to complex activities like ironing a shirt.
These interactions reflect the complex nature of object affordances \cite{gibson2014ecological}, which are essential for understanding and synthesizing realistic 3D environments.
Modeling these dynamics between humans and objects is crucial for many downstream applications in computer vision and graphics, such as character animation, immersive VR/AR, robotics, and product design.
In this work, we focus on generating diverse and realistic 4D HOI motions from easy-to-use text prompts beyond a limited taxonomy of interactions.

While humans instinctively recognize how to interact with objects, replicating this behavior in machines requires careful \emph{planning} and \emph{joint reasoning} of affordances, body motions, and object movements.
Prior works~\cite{diller2024cg,peng2025hoidiff,li2024controllable,li2024zerohsi,kim2025david} typically model interactions as overall whole-body and object motions without explicitly reasoning about the underlying \emph{part-level} mechanics.
However, HOI is not merely global proximity of a person to an object but rather a coordinated engagement between specific body parts and functional object parts, which we term \emph{part-level affordances}.

We present \OURS{}, a zero-shot framework that prioritizes this part-level reasoning to generate realistic 4D HOI motions.
Key to our approach is an explicit planning stage to define how specific object parts relate to human body parts before motion synthesis.
We leverage the emergent reasoning capabilities of large language models (LLMs) \cite{guo2025deepseek} to imagine the part-level mechanics of these interactions.
We then formalize the reasoning result into a structured representation called \emph{part affordance graph} (PAG), where nodes correspond to object or body parts, and edges encode their contact relations.

Given as input a set of 3D objects and a text prompt describing the desired interaction, \OURS{} generates the corresponding human and object 4D motion sequences.
We use the PAG as a high-level interaction scaffolding to guide the distillation of 4D motions from video diffusion models \cite{yang2024cogvideox} in a zero-shot fashion.
Concretely, the PAG 
(1) is grounded to 3D object geometry for semantic part segmentation required for the interaction;
(2) informs video diffusion to generate a reference video adhering to the interaction plan;
and (3) serves as contact constraints in a part affordance-guided optimization to lift the reference video into 4D HOI motions.

\OURS{} offers a general formulation that extends beyond the single-person, single-object scenarios tackled by the state-of-the-art~\cite{peng2025hoidiff,li2024controllable}.
The flexibility of PAGs allows for generating complex interactions involving multiple people and multiple objects (\cref{fig_teaser}) simply by expanding the graph nodes and edges to reflect new affordances.
We demonstrate the effectiveness of our approach through extensive experiments on a variety of interaction scenarios, including single and multi-person/object interactions.
Perceptual studies show that our method significantly outperforms state-of-the-art methods \cite{peng2025hoidiff,li2024controllable} in terms of interaction realism and alignment with text prompts.

The contributions of our work\footnote{Project page: \href{https://craigleili.github.io/projects/hoipage/}{craigleili.github.io/projects/hoipage}} are summarized as follows:
\begin{enumerate}[leftmargin=*, labelwidth=1.5em, labelsep=0.5em, topsep=-0.5em, itemsep=0em, parsep=0em]
  \item We introduce the first zero-shot 4D HOI synthesis approach that explicitly prioritizes part-level reasoning. We propose part affordance graphs, a structured representation serving as universal scaffolding to ground the synthesis in part-level interaction mechanics.
  \item We formulate a part affordance-guided optimization to distill 4D HOIs from video diffusion, achieving realistic part-level contact in generated human-object trajectories.
  \item Our part affordance formulation is flexible and versatile, enabling generalization to diverse interaction scenarios, including multi-person/object interactions.
\end{enumerate}
\vspace{-0.1cm}

\section{Related Work}
\label{sec_related_work}

\mypara{Human Motion Generation.}
4D human motion synthesis has seen significant advances in recent years, largely driven by advances in deep learning.
Earlier work leveraged recurrent neural networks for synthesis \cite{fragkiadaki2015recurrent,aksan2019structured,gopalakrishnan2019neural,martinez2017human}.
More recently, with the success of denoising diffusion models \cite{sohl2015deep,song2020denoising,ho2020denoising}, diffusion-based human motion generation has become a powerful and widely adopted approach to synthesizing human motion \cite{zhang2023tedi,raab2023single,zhao2023modiff,dabral2023mofusion,tevet2023human,shafir2023human,zhang2022motiondiffuse,karunratanakul2024optimizing,jiang2023motiongpt,petrovich2024multi}.
These methods show remarkable motion synthesis results, but focus on modeling human motion in isolation, without interactions intrinsic to real-world scenarios.
 
\mypara{Human-Object Interaction Generation.}
As interactions play a crucial role in 4D synthesis, various approaches have focused on modeling HOIs, generating the motion of a single human and single object.
Several works tackled this task under the assumption of a static object \cite{taheri2022goal,tendulkar2023flex,zhang2022couch,wu2022saga,lee2023locomotion,zhang2023roam,kulkarni2023nifty}, focusing only on human motion generation.
Recently, new methods have proposed to generate both human and object motion for single-person single-object scenarios \cite{li2023object,wan2022learn,diller2024cg,peng2025hoidiff,li2024controllable,wu2024thor,xu2024interdreamer,xu2023interdiff,wang2023physhoi,yang2024f} and multi-object scenarios \cite{lv2024himo}.
A parallel line of work targets dexterous hand-object interaction synthesis~\cite{zhang2026openhoi,han2025touch}, complementary to the full-body interaction setting.
These methods can synthesize realistic HOIs, but rely on ground truth real-world captures of HOI data to train the generative models. Collecting such 4D ground truth data is very time-consuming and expensive, and thus limited in size and diversity \cite{bhatnagar22behave,taheri2020grab,jiang2023full}.
In contrast, our approach proposes a general approach to handle various novel, diverse objects without requiring any 4D interaction data for training.

GenZI~\cite{li2024genzi} recently introduced a new paradigm for 3D human-scene interaction synthesis, by distilling priors from text-to-image foundation models to generate interactions without requiring 3D interaction training data, focusing only on static interaction generation \cite{zhang2025interactanything,zhu2024dreamhoi,yang2024lemon,kim2024beyond}. 
Concurrent to our approach, ZeroHSI~\cite{li2024zerohsi}, DAViD~\cite{kim2025david}, and ZeroHOI~\cite{lou2025zero} have begun to address the challenge of zero-shot 4D HOI synthesis to circumvent the need for 4D ground truth training data. 
While these approaches also leverage knowledge from large video foundation models, they treat the human-object motion globally, lacking finer-grained interaction modeling at the level of parts. This limits the ability to capture complex contact dynamics and multi-object or multi-person interactions. For instance, ZeroHOI~\cite{lou2025zero} targets single-person single-object scenarios without an explicit part-level structure. In contrast, our approach introduces a structured Part Affordance Graph that explicitly encodes part-level contact and relative motion relations, which naturally extends to multi-interaction scenarios.

\begin{figure*}[t]
    \centering
    \includegraphics[width=\textwidth]{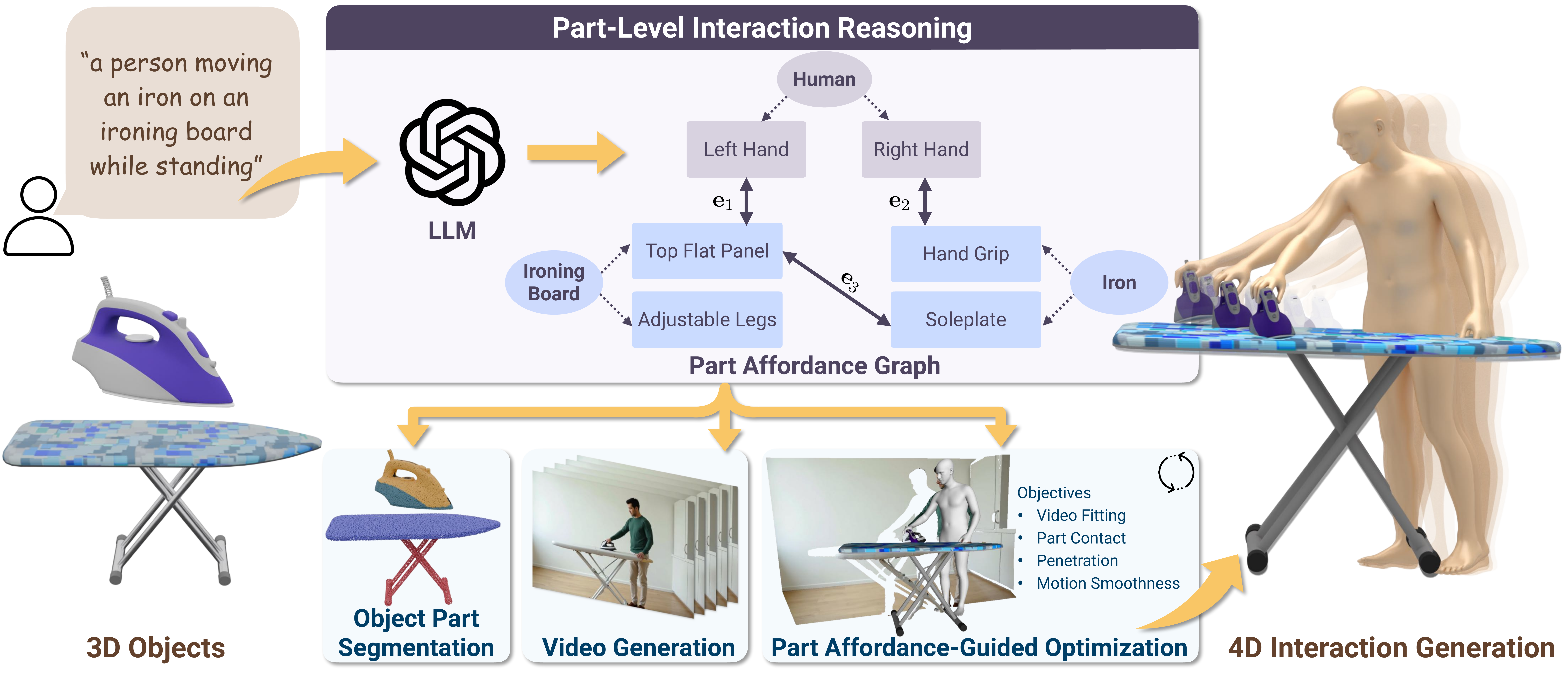}
    \caption{\OURS{} generates realistic 4D human-object interaction (HOI) motions from a set of 3D objects and a text prompt. We introduce a part-level interaction reasoning stage (\textbf{top-middle}), leveraging a large language model (LLM) to imagine how specific object parts relate to human body parts. The reasoning is captured by a part affordance graph (PAG), serving as a high-level interaction scaffolding to guide the synthesis process (\textbf{bottom-middle}): 3D object part segmentation, HOI video generation, and 4D HOI optimization.
    }
    \label{fig_pipeline}
\end{figure*}

\mypara{3D Affordance Analysis.}
Various works have also proposed to study 3D affordances via structured graph representations to capture relations between humans and objects.
PiGraphs~\cite{savva2016pigraphs} introduced a prototypical interaction graph representation to capture physical contact and visual attention relations between human body parts and 3D scenes, in order to synthesize static snapshots of human-scene interactions. 
In contrast to the graph-based representation, Fisher et al.~\cite{fisher2015activity} propose an activity heatmap representation learned from human-scene interactions for synthesizing new 3D scenes that enable similar interactions. 
iMapper~\cite{monszpart2019imapper} instead proposes to leverage ``scenelets'' that capture short interaction subsequences as a database prior to reconstruct a human and the objects interacted with from monocular video observations.
Inspired by these methods, we also propose to explicitly model affordance relations, as part-based affordance graphs of (multi-) human-object interactions for zero-shot 4D human-object interaction synthesis.

\section{Method}
\label{sec_method}

We aim to generate realistic 4D motion sequences of human-object interactions conditioned on text descriptions in a zero-shot manner.
Our approach, \OURS{}, introduces an explicit planning stage to first imagine part-level interaction mechanics before motion synthesis.
This planning is enabled by LLMs to perform holistic reasoning over the part-level affordances, ensuring a more grounded relation between humans and objects during interactions. We capture the reasoning into a part affordance graph representation, which serves as a structural scaffolding for the entire generation process.
The flexibility of PAGs allows for synthesizing diverse, complex HOI scenarios (\cref{fig_teaser}), including (1) single-person single-object, (2) multi-person single-object, and (3) single-person multi-object interactions.
Our approach is illustrated in \cref{fig_pipeline}.

For a given set of 3D objects $\{\Object\}$ and a short text prompt $\TextPrompt$ describing the desired 4D interaction, \OURS{} produces a sequence of poses $\{ (\ObjectRotation_{t}, \ObjectTranslation_{t}) \}_{t=1}^{\MotionLength}$ for each object $\Object$ and a sequence of body parameters $\{ \HumanParameters_{t} \}_{t=1}^{\MotionLength}$ for each human $\Human$, where $\MotionLength$ is the number of frames.
Object $\Object$ is a textured 3D mesh, and human $\Human$ is a SMPL-X body model~\cite{pavlakos2019expressive}.
At time $t$, each object pose is represented by a 3D rotation $\ObjectRotation_{t}$ and a 3D translation $\ObjectTranslation_{t}$, while $\HumanParameters_{t}$ includes body joint rotations, body shape coefficients, a global rotation, and a global translation.
We omit the indexing of objects and humans for simple notation.

\subsection{Interaction Planning with Part Affordance Graphs}
\label{subsec_method_part_affordance_graphs}

Our first step is to develop a high-level plan for the desired interaction by reasoning about how human body parts and object parts should relate to each other conceptually during the interaction. This reasoning determines part semantics, contact, and motion dynamics constraints to be grounded in the subsequent interaction motion synthesis.

We formulate the interaction plan as a part affordance graph $\Graph = (\GraphNodes, \GraphEdges)$, with nodes $\GraphNodes$ and edges $\GraphEdges$ (\cref{fig_pipeline} top-middle).
Each node $\GraphNode \in \GraphNodes=\GraphNodes_o\cup\GraphNodes_h$ corresponds to a part from either the object ($\GraphNodes_o$) or the body ($\GraphNodes_h$) part collection.
To represent a whole object or human, we also add a virtual parent node $\GraphVirtualNode$ to $\GraphNodes$, connected to all its constituent part nodes.
For each object $\Object$, this virtual parent node $\GraphVirtualNode_o$ has two motion attributes $(\GraphNodeRotational, \GraphNodeTranslational)$, indicating if the object rotates ($\GraphNodeRotational$) or translates ($\GraphNodeTranslational$).
If both states are false, the object remains stationary throughout the interaction.

Each edge $\GraphEdge \in \GraphEdges$ in the graph represents the contact of an object part to a human body part, or to another object part.
Each edge $\GraphEdge$ has two attributes $(\GraphEdgeContinuous, \GraphEdgeStatic)$:
$\GraphEdgeContinuous$ indicates if the contact is continuous throughout the interaction, while $\GraphEdgeStatic$ denotes if the contact is relatively static.
For example, in \cref{fig_pipeline}, the edge $\GraphEdge_2$ represents the contact between the right hand and the iron's hand grip as continuous and relatively static ($\GraphEdgeContinuous=\text{true}, \GraphEdgeStatic=\text{true}$).
The edge $\GraphEdge_3$ between the ironing board's top flat panel and the iron's soleplate is described as continuous but not static.
PAGs are flexible and can represent complex scenarios involving multiple people or objects by simply adding more nodes and edges.

We leverage an LLM~\cite{guo2025deepseek} to plan and infer the PAG $\Graph$ from the text prompt $\TextPrompt$. The LLM needs to identify the required object parts, the number of people involved, and the part contact edges. We use a pre-defined set of 12 human body parts (e.g., left/right hand, left/right foot, hips). 
LLMs are well-suited for this task because they can reason about the common ways humans interact with various objects based on their extensive knowledge base and powerful in-context learning capabilities.
While vision-language models (VLMs) could be used by prompting them with both text descriptions and rendered object images, we found that the VLMs we experimented with occasionally ignore the visual input, partly due to the known hallucination issue~\cite{liu2024survey}, and they are less robust to infer plausible PAGs.
We thus opt for LLMs but stress that our PAG representation is agnostic to the foundation model used, and VLMs could be used alternatively as they continue to improve.

The resulting PAG serves as a universal scaffolding that instructs all subsequent generation stages. The inferred part nodes and contact constraints are grounded in 3D objects for part segmentation, used to guide video generation, and finally enforced during 4D HOI lifting optimization.

\subsection{Grounding Abstract Parts to 3D Geometry}
\label{subsec_method_multiview_object_part_segmentation}

Once the interaction plan is established, we first ground the abstract object part nodes $\GraphNodes_o$ from the PAG $\Graph$ to the actual 3D geometry of each object $\Object$. This leads to fine-grained semantic segmentation of the input 3D objects, facilitating the realization of part affordances.
We first render $\Object$ from 8 sampled virtual views.
Open-vocabulary detection~\cite{bai2025qwen2} is performed on the rendered images to obtain each object part's bounding box, and then we predict 2D part masks within these boxes~\cite{ravi2024sam2}. These part masks are aggregated back into 3D through voting on the point cloud sampled from the object.

\subsection{Grounding Interaction Dynamics in Video}
\label{subsec_method_hoi_video_generation}

To bridge the gap between abstract planning and 4D motions, we embed the PAG $\Graph$ into a reference HOI video depicting the planned affordance dynamics. This temporal sequence provides rich motion cues for 4D interaction generation. 

\mypara{Part Affordance-Guided Video Generation.}
We generate the interaction video $\{\VideoFrame_{t}\}_{t=1}^{\MotionLength}$, where $\VideoFrame_{t}$ is the frame at time $t$, using video diffusion~\cite{yang2024cogvideox}.
To inform video generation with the imagined interaction, we translate the part-level contact and motion states from the PAG $\Graph$ into a more detailed video description $\TextPrompt^+$ using the LLM~\cite{guo2025deepseek} conditioned on the original text prompt $\TextPrompt$.

\begin{figure}[t]
    \centering
    \includegraphics[width=\columnwidth]{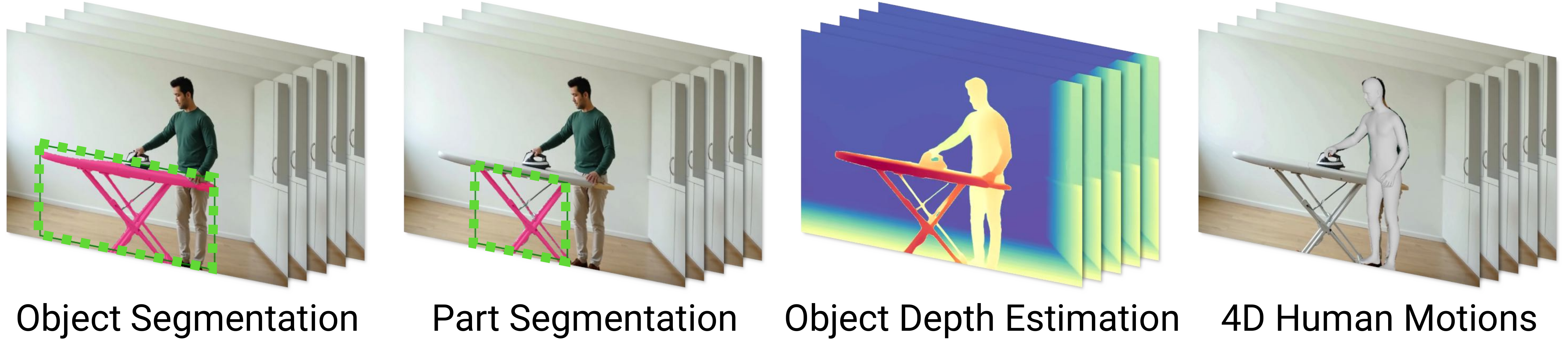}
    \caption{Inferred object constraints and human motions from a generated interaction video.
    \vspace{-0.3cm}
    }
    \label{fig_video_processing}
\end{figure}

\mypara{Extracting Video Constraints.}
From the generated video, we extract a rich set of constraints, including part-level 2D-3D object correspondence, video object geometry, and human poses, for 4D interaction lifting optimization.

We detect, track, and segment~\cite{bai2025qwen2,ravi2024sam2} each object and its constituent parts across the video frames using the part nodes defined in the PAG $\Graph$.
This produces a sequence of object masks for each object node $\GraphVirtualNode_o$, and part masks for each object part node $\GraphNode \in \GraphNodes_o$, as shown in \cref{fig_video_processing}-left.

Depth (\cref{fig_video_processing}-middle) is estimated for each frame~\cite{wang2024moge}, and we combine it with the above video segmentation masks to back-project the video into a sequence of 3D point clouds for each object and its parts.

We also perform 4D human motion recovery~\cite{shen2024gvhmr} on the generated video to extract body parameters $\{\HumanParameters_t\}_{t=1}^{\MotionLength}$ for each person.
However, this only estimates human motions in isolation. The final optimization (\cref{subsec_method_hoi_optimization}) will reconcile these motions with the part-level affordance constraints to achieve the interaction dynamics outlined in the PAG $\Graph$.

\subsection{Part Affordance-Guided 4D HOI Optimization}
\label{subsec_method_hoi_optimization}
The final stage is a part affordance-guided optimization that lifts the reference video into a realistic 4D interaction.
We optimize for the object motion sequences $\{ (\ObjectRotation_{t}, \ObjectTranslation_{t}) \}_{t=1}^{\MotionLength}$ based on the PAG $\Graph$ to achieve plausible part-level relations between the 3D objects $\{\Object\}$ and the recovered human bodies $\{\HumanParameters_t\}_{t=1}^{\MotionLength}$.
The objectives include that objects fit well to the generated video at the part level, object motions respect the part contact in $\Graph$ while avoiding penetration, and the resulting object motions are temporally smooth.

\mypara{Part-Based Fitting.}
To ensure the 3D objects $\{\Object\}$ follow the object motions in the video, we define a fitting loss $\LossFit$ that incorporates part-level alignment in both 3D and 2D. 

Specifically, in 3D at time $t$, we compute the Chamfer Distance $\LossFit^{\text{3D}}$ between each 3D part point cloud (\cref{subsec_method_multiview_object_part_segmentation}) of object $\Object$, transformed by $(\ObjectRotation_{t}, \ObjectTranslation_{t})$, and its corresponding 3D part point cloud derived from the video frame depth (\cref{subsec_method_hoi_video_generation}).
In 2D, we project the 3D part point clouds, again transformed by $(\ObjectRotation_{t}, \ObjectTranslation_{t})$, onto the image plane using the estimated camera intrinsics of the generated video. Then the Chamfer Distance $\LossFit^{\text{2D}}$ is computed between the projected object part point clouds and the corresponding 2D part mask pixels from the video.

Similarly, we also compute the fitting loss terms, $\LossFit^{\text{3D}}$ and $\LossFit^{\text{2D}}$, at the object level in both 3D and 2D.
The object-level fitting helps to mitigate any effect from potentially inaccurate part segmentations, while the part-level fitting loss can help to find better correspondence between the 3D objects and the generated video.
Overall, the fitting loss is $\LossFit = \LossFit^{\text{3D}} + \LossFit^{\text{2D}}$.

\mypara{Part-Level Contact.}
We compute the contact loss on a part basis guided by each edge $\GraphEdge = (\GraphNode_1, \GraphNode_2)$ and its attribute $\GraphEdgeContinuous$ (continuous vs. non-continuous) from the PAG $\Graph$:
\begin{equation}
    \label{loss_contact_continuity}
	\scalebox{0.8}{$
    \LossContactContinuity = \sum_{\GraphEdge = (\GraphNode_1, \GraphNode_2) \in \GraphEdges}
    \begin{cases}
        \frac{1}{\MotionLength} \sum_{t=1}^{\MotionLength} \text{MD}(\GraphEdgeNodePointCloud^{\GraphNode_1}_t, \GraphEdgeNodePointCloud^{\GraphNode_2}_t), & \text{if} \ \GraphEdgeContinuous = \text{true}\\
        \min_t \text{MD}(\GraphEdgeNodePointCloud^{\GraphNode_1}_t, \GraphEdgeNodePointCloud^{\GraphNode_2}_t), & \text{otherwise}
    \end{cases}
	$}
\end{equation}
where $\GraphEdgeNodePointCloud^{\GraphNode_1}_t$ and $\GraphEdgeNodePointCloud^{\GraphNode_2}_t$ are the 3D part point clouds of $\GraphNode_1, \GraphNode_2$ at time step $t$, respectively, and can be either a 3D object part or a human body part.
$\text{MD}(\cdot)$ denotes the minimum distance among any pair of nearest neighbors between two point clouds.
The top case is for continuous contact across the $\MotionLength$ frames, while the bottom is for non-continuous contact.

We also measure relative contact dynamics between $\GraphNode_1, \GraphNode_2$ based on the attribute $\GraphEdgeStatic$ (relatively static vs. dynamic) of each graph edge $\GraphEdge$:
\begin{equation}
    \label{loss_contact_dynamics}
    \scalebox{0.75}{$
    \LossContactDynamics = \sum\limits_{\GraphEdge = (\GraphNode_1, \GraphNode_2) \in \GraphEdges} \sum_{t}
    \begin{cases}
         \LossEuclidean(\GraphEdgeNodePointCloud^{\GraphNode_2 \rightarrow \GraphNode_1}_{t}, \GraphEdgeNodePointCloud^{\GraphNode_2 \rightarrow \GraphNode_1}_{t+1}), & \text{if} \ \GraphEdgeStatic = \text{true}\\
         \LossEuclidean(\GraphEdgeNodePointCloud^{\GraphNode_2 \rightarrow \GraphNode_1}_{t}, \frac{1}{2} (\GraphEdgeNodePointCloud^{\GraphNode_2 \rightarrow \GraphNode_1}_{t-1} + \GraphEdgeNodePointCloud^{\GraphNode_2 \rightarrow \GraphNode_1}_{t+1})), & \text{otherwise}
    \end{cases}
    $}
\end{equation}
where $\GraphEdgeNodePointCloud^{\GraphNode_2 \rightarrow \GraphNode_1}_{t}$ denotes the 3D part point cloud of node $\GraphNode_2$ at time step $t$ transformed to the canonical object space of node $\GraphNode_1$ by the inverse object pose $(\ObjectRotation_{t}, \ObjectTranslation_{t})$ of $\GraphNode_1$, assuming $\GraphNode_1$ is always an object part node.
$\LossEuclidean (\cdot)$ measures the average Euclidean distance of each corresponding point pairs in two point clouds.
The top case measures static contact, while the bottom promotes dynamic but temporally coherent contact.
Overall, the contact loss is $\LossContact = \LossContactContinuity + \LossContactDynamics$.

\mypara{Penetration.}
We compute a penetration loss $\LossPenetration$ for all object-human pairs.
A signed distance field is pre-computed for each 3D object for measuring the penetration depth between vertices of a human body and the object surface. This follows established practice in human-object penetration loss for interactions \cite{li2024genzi,Hassan_2019_ICCV}.

\mypara{Temporal Smoothness.}
We regularize the object motions $\{ (\ObjectRotation_{t}, \ObjectTranslation_{t}) \}_{t=1}^{\MotionLength}$ to be temporally smooth based on the motion state attributes $(\GraphNodeRotational, \GraphNodeTranslational)$ of each virtual object node:
\begin{equation}
    \label{loss_temporal_smoothness_rotation}
	\scalebox{0.9}{$
	\LossSmoothnessRotation = \sum_{\Object} \sum_{t}
	\begin{cases}
		\text{GD} (\ObjectRotation_{t}, \frac{1}{2}(\ObjectRotation_{t-1} + \ObjectRotation_{t+1})), & \text{if} \ \GraphNodeRotational = \text{true}\\
	    \text{GD} (\ObjectRotation_{t}, \ObjectRotation_{t+1}), & \text{otherwise}
	\end{cases}
	$}
\end{equation}
where $\text{GD}(\cdot)$ denotes the geodesic distance between two rotations.
The top case, where spherical linear interpolation is used, promotes smooth rotational motions for object $\Object$, while the bottom penalizes temporal changes in object rotations.
For the translations, we compute
\begin{equation}
    \label{loss_temporal_smoothness_translation}
	\scalebox{0.9}{$
	\LossSmoothnessTranslation = \sum_{\Object} \sum_{t}
	\begin{cases}
		\LossEuclidean(\ObjectTranslation_{t}, \frac{1}{2} ( \ObjectTranslation_{t-1} +  \ObjectTranslation_{t+1})), & \text{if} \ \GraphNodeTranslational = \text{true}\\
	    \LossEuclidean(\ObjectTranslation_{t}, \ObjectTranslation_{t+1}), & \text{otherwise}
	\end{cases}
	$}
\end{equation}
where the top case promotes smooth translational motions for object $\Object$, while the bottom penalizes temporal changes in object translations.
Overall, the temporal smoothness loss is $\LossSmoothness = \LossSmoothnessRotation + \LossSmoothnessTranslation$.

\mypara{Total Loss.}
Our total loss is a weighted sum of the fitting, contact, penetration, and temporal smoothness terms:
$\LossTotal = \LossFitWeight \LossFit + \LossContactWeight \LossContact + \LossPenetrationWeight \LossPenetration + \LossSmoothnessWeight \LossSmoothness$.

\subsection{Implementation Details}
\label{subsec_method_implementation}

Our \OURS{} is implemented using PyTorch~\cite{Pytorch:NIPS2019}.
To improve realism of a synthesized HOI video (\cref{subsec_method_hoi_video_generation}),
we generate 5 candidate images for the first frame using FLUX \cite{fluxtext2image} and then select the one with the best visual quality \wrt{} human anatomy, text alignment, and camera views by querying a VLM (GPT-4.1).
We use 50 denoising steps for both image and video diffusion.
CogVideoX generates 49 frames per video, and thus $\MotionLength=49$.
We optimize $\LossTotal$ for 600 steps using gradient descent, which takes $\sim$6 mins for single-object interactions and $\sim$10 mins for interactions involving 2 objects on A100 GPUs.
We repeat the optimization for 4 times with different sampled object rotation initializations around the up axis to mitigate convergence to local optimum caused by Chamfer Distance in $\LossFit$.
Prompts for part affordance graph inference with LLMs and first-frame selection with VLMs are provided in the appendix.

\section{Experiments}
\label{sec_experiments}

\begin{figure*}[h!]
    \centering
    \includegraphics[width=0.98\textwidth]{figures/sketchfab_qualitative_comparisons_single_interaction_.pdf}
    \caption{Single-person single-object interaction generations on the Sketchfab dataset. Our part affordance-guided approach generates more realistic 3D interaction motions with better text prompt alignment, compared to the baselines HOI-Diff and CHOIS, which struggle to generalize across diverse 3D objects (\eg, lawnmower) unseen during training.
    }
    \label{fig_sketchfab_qualitative_comparisons_single_interaction}
\end{figure*}

\begin{table*}[t]
	\centering
	\caption{Comparing single-person single-object interaction generations on the Sketchfab dataset. Our part affordance-guided approach generates realistic human-object interaction motions with semantic consistency, temporal smoothness, motion diversity, and physical plausibility metrics outperforming the baselines HOI-Diff and CHOIS that require 4D interaction data for supervision.
	}
	\resizebox{0.7\textwidth}{!}{%
		\begin{tabular}{l|c|cc|cc|cc}
			\hline
			         & Semantics            & \multicolumn{2}{c|}{Temporal Smoothness} & \multicolumn{2}{c|}{Motion Diversity} & \multicolumn{2}{c}{Physical Plausibility}                                                                     \\
			         & VideoCLIP $\uparrow$ & Human $\downarrow$                       & Object $\downarrow$                   & Human $\uparrow$                          & Object $\uparrow$ & Non-collision $\uparrow$ & Contact $\uparrow$ \\
			\hline
			HOI-Diff & 0.233                & \textbf{0.007}                           & 0.035                                 & 0.35                                      & 0.72              & 0.98                     & 0.76               \\
			CHOIS    & 0.239                & 0.009                                    & 0.009                                 & 0.44                                      & 0.49              & 0.98                     & 0.64               \\
			Ours     & \textbf{0.250}       & 0.008                                    & \textbf{0.006}                        & \textbf{0.47}                             & \textbf{0.80}     & \textbf{0.99}            & \textbf{0.92}      \\
			\hline
		\end{tabular}
	}
	\label{tab_sketchfab_quantitative_comparisons_single_interaction}
\end{table*}

\begin{table*}[t]
	\centering
	\caption{Multi-person single-object (MPSO) and single-person multi-object (SPMO) interaction generations on the Sketchfab dataset. Our approach handles well complex interaction scenarios involving multiple persons/objects, owing to the flexibility of our part affordance graphs, while achieving consistent performance in the perceptual ratings (on a scale of 1-5) and evaluation metrics.
	}
	\resizebox{0.8\textwidth}{!}{%
		\begin{tabular}{l|cc|c|cc|cc|cc}
			\hline
			     & \multicolumn{2}{c|}{Perceptual} & Semantics            & \multicolumn{2}{c|}{Temporal Smoothness} & \multicolumn{2}{c|}{Motion Diversity} & \multicolumn{2}{c}{Physical Plausibility}                                                                                        \\
			     & Realism$\uparrow$               & Text Match$\uparrow$ & VideoCLIP $\uparrow$                     & Human $\downarrow$                    & Object $\downarrow$                       & Human $\uparrow$ & Object $\uparrow$ & Non-collision $\uparrow$ & Contact $\uparrow$ \\
			\hline
			MPSO & 4.17                            & 4.46                 & 0.312                                    & 0.009                                 & 0.002                                     & 0.43             & 0.79              & 0.99                     & 0.62               \\
			SPMO & 4.46                            & 4.59                 & 0.268                                    & 0.005                                 & 0.005                                     & 0.54             & 0.87              & 0.99                     & 0.90               \\
			\hline
		\end{tabular}
	}
	\vspace{-0.2cm}
	\label{tab_sketchfab_quantitative_comparisons_multi_interaction}
\end{table*}

\begin{figure*}[t]
    \centering
    \includegraphics[width=\textwidth]{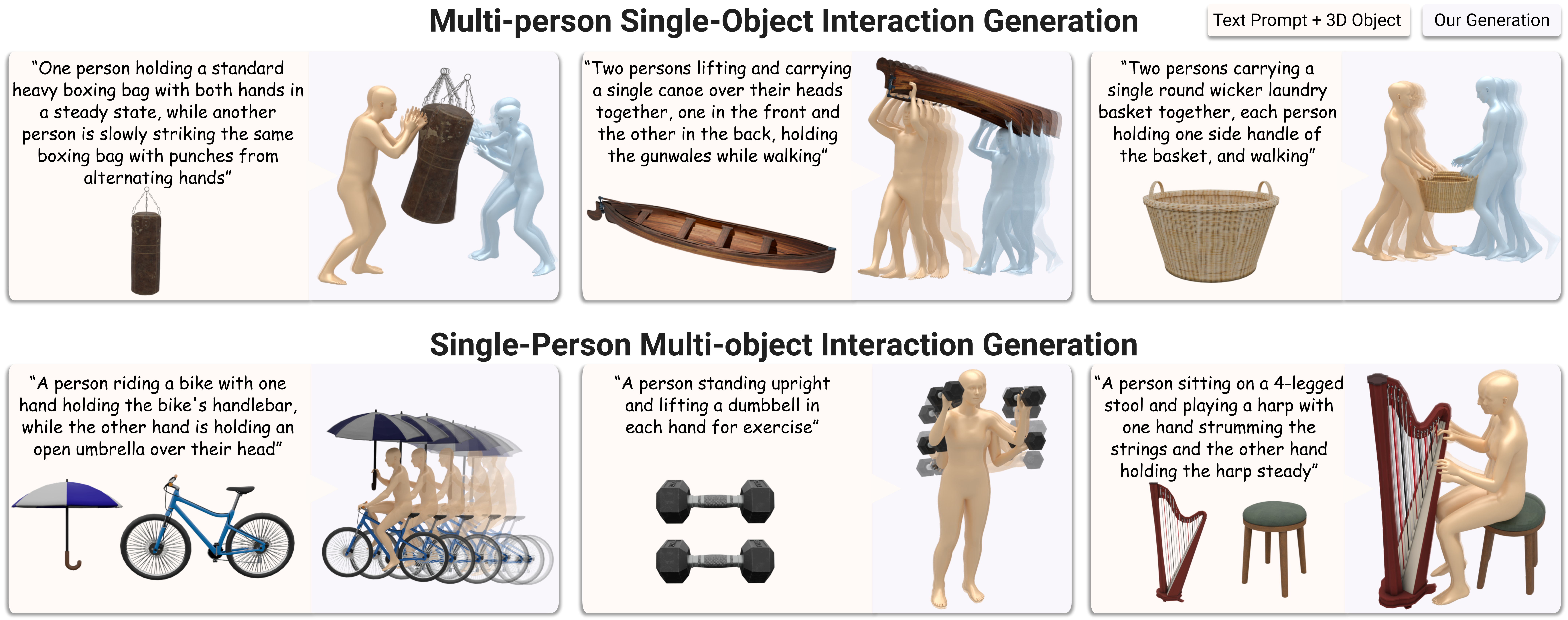}
    \caption{Our multi-person single-object and single-person multi-object interaction generations on the Sketchfab dataset. The flexibility of part affordable graphs enables our approach to generate diverse 3D interactions with multiple persons/objects.
    \vspace{-0.2cm}
    }
    \label{fig_sketchfab_qualitative_comparisons_multi_interaction}
\end{figure*}

\begin{figure}[t]
    \centering
    \includegraphics[width=\columnwidth]{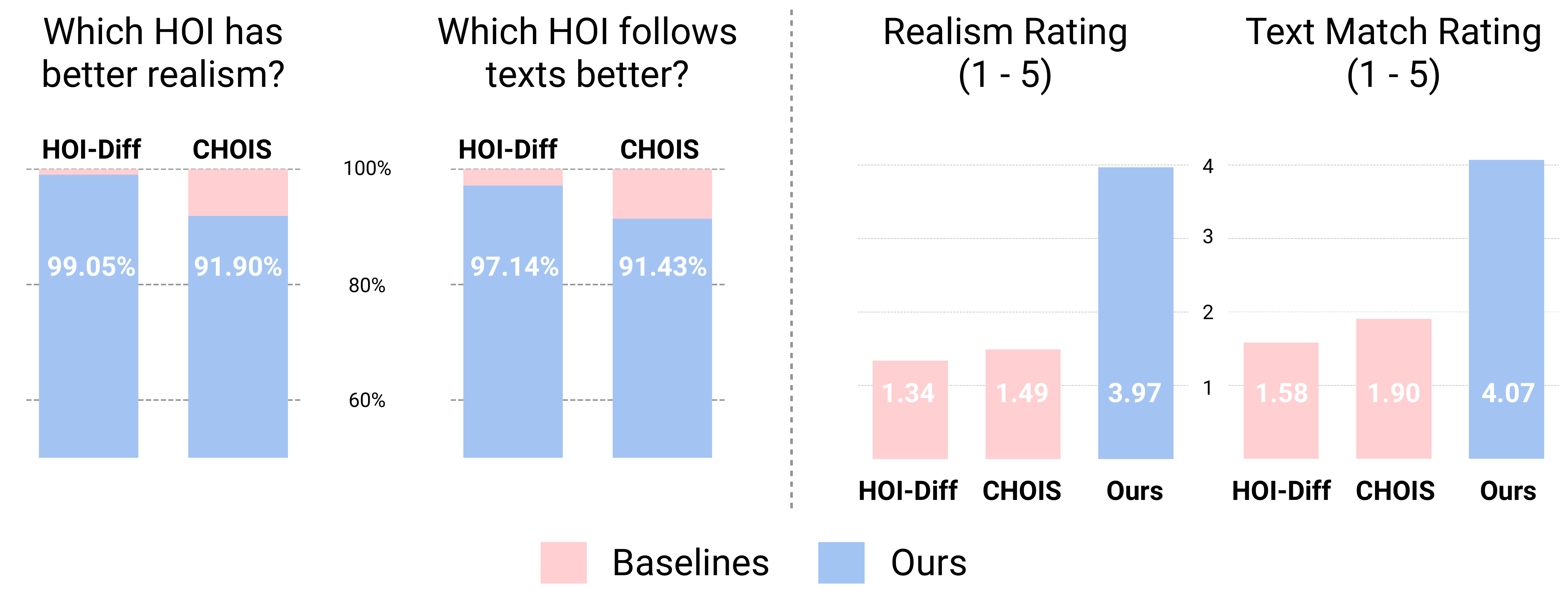}
    \caption{Perceptual studies of single-person single-object interaction generations on the Sketchfab dataset. In the binary study (\textbf{left}), participants strongly prefer our method over the baselines HOI-Diff and CHOIS for interaction realism and text matching. In the unary study (\textbf{right}), our generations achieve the highest ratings (on a scale of 1-5) compared to the baselines.
    \vspace{-0.5cm}
    }
    \label{fig_sketchfab_perceptual_study}
\end{figure}

We evaluate \OURS{} both qualitatively and quantitatively in diverse interaction scenarios, including single-person single-object, multi-person single-object, and single-person multi-object interactions.
Our approach achieves superior generation realism, diversity, and text alignment when compared to the state-of-the-art methods \cite{peng2025hoidiff,li2024controllable}. %

\mypara{Dataset.}
We collected 24 daily objects from Sketchfab.com, spanning categories such as household items, sports equipment, instruments, and transportation devices.
Each object is a textured 3D mesh and canonicalized with a consistent upright orientation.
A signed distance field (SDF) is precomputed for each object.
We prepared 16 text prompts for single-person single-object interactions and 5 prompts for multi-person or multi-object scenarios, respectively.

\mypara{Baselines.}
We compare with the state-of-the-art HOI-Diff \cite{peng2025hoidiff} and CHOIS \cite{li2024controllable}, which generate \textit{single-person single-object interactions} from text prompts.
These baselines were trained on real-world captured data of people interacting with indoor objects.
We use the pre-trained models released by the authors and adapt them to the Sketchfab dataset, as we do not have any 4D ground truth for this data for training.
CHOIS additionally requires object waypoints as input, which we provide by using the object waypoints generated by our approach.

\mypara{Evaluation Metrics.}
\newline
\emph{- Perceptual Study.}
We evaluate the realism and text alignment of 4D HOI motions.
In a binary study, participants are shown two rendered interaction videos and asked to select the more realistic one and the one better matching a given text prompt, respectively.
In a unary study, they are shown a single interaction video and asked to rate its realism and text alignment, respectively, from 1 (= strongly disagree) to 5 (= strongly agree).
We surveyed 30 participants.\\
\emph{- Semantic Alignment.}
To measure alignment between a 4D HOI and a text prompt, we compute the cosine similarity between the text and the rendered video embeddings.
A pre-trained VideoCLIP model~\cite{bolya2025perception} (PE-Core-G14-448) is used to extract the embeddings.
We render a 4D interaction from 3 different views and compute the average cosine similarity as the score.\\
\emph{- Temporal Smoothness.}
We evaluate the temporal smoothness of a generated 4D human motion by computing the distance between each 3D joint position at a given frame and the average position of the same joint in the two neighboring frames (similar to \cref{loss_temporal_smoothness_translation}-top).
Similarly, the temporal smoothness of a 4D object motion is computed using the object's bounding box corners.\\
\emph{- Motion Diversity.}
To evaluate human motion diversity, we generate 5 interaction samples for each text prompt and compute the distance between each pair of samples for every joint position at a given frame.
Object motion diversity is evaluated in the same way \wrt{} bounding box corners.\\
\emph{- Physical Plausibility (Non-collision, Contact).}
We compute non-collision and contact scores of a generated 4D interaction.
At each frame, we check for collisions by querying each object's SDF for all human body vertices~\cite{zhao2022compositional}.
The non-collision score is defined as the ratio of the number of non-colliding human body vertices to the total number of vertices at each frame.
The contact score is computed as the ratio of the number of frames with collision to the sequence length.

\subsection{Comparison to Baselines}

\mypara{Quantitative Evaluation.}
The perceptual study results are shown in \cref{fig_sketchfab_perceptual_study}.
In the binary evaluation, our 4D interaction generations are strongly preferred over HOI-Diff and CHOIS, receiving more than 91\% of the votes for both realism and text alignment.
In the unary evaluation, 
participants rated our generations with an average score of $\sim$4 for both criteria, significantly higher than HOI-Diff and CHOIS, which scored below 2.
In \cref{tab_sketchfab_quantitative_comparisons_single_interaction}, our approach achieves the best scores in semantic alignment, temporal smoothness of object motions, motion diversity, and physical plausibility metrics.
HOI-Diff has slightly better temporal smoothness for human motions, but its generations do not align well with the text prompts and have the lowest human motion diversity.
In contrast, our approach generates more diverse human motions.
The perceptual studies and quantitative results show that our part-level contact distillation from LLMs is effective in generating more realistic and text-aligned interactions.

\mypara{Qualitative Evaluation.}
\cref{fig_sketchfab_qualitative_comparisons_single_interaction} presents comparisons of generated 4D interactions.
HOI-Diff and CHOIS struggle to generate plausible interactions for the Sketchfab objects unseen during their training.
For example, HOI-Diff produces nearly static human poses with the guitar and has significant penetration with the lawnmower, while CHOIS generates less precise part-level contact between human hands and the lawnmower handle. 
In contrast, our approach generalizes better across different objects in zero shot, capturing well part-level affordances of objects.

\subsection{Multi-interaction Evaluation}
In contrast to fully-supervised baselines that require real-world 4D captures for training \cite{peng2025hoidiff,li2024controllable}, our zero-shot part-guided approach enables synthesizing more general, complex interaction scenarios, such as multi-person single-object generation and single-person multi-object generation.
\cref{fig_sketchfab_qualitative_comparisons_multi_interaction} shows our approach on these multi-interaction scenarios, by simply distilling multi-person or multi-object nodes and their corresponding part nodes from the LLM during PAG construction.
We also quantitatively evaluate our multi-interaction generation in \cref{tab_sketchfab_quantitative_comparisons_multi_interaction}.
Although contact can become more challenging with the multi-person scenario, with more human contact constraints to satisfy, our approach synthesizes interaction sequences of quality that closely matches the simpler single-person single-object interactions in these more complex interaction scenarios.
More results on multi-person multi-object generation are provided in the appendix.
\begin{table}[t]
	\centering
	\caption{Ablation studies on Sketchfab. Results are averaged over multi-person single-object and single-person multi-object interaction generations. Object motion smoothness, diversity, and physical contact scores degrade significantly without part-level fitting (PF), part-level contact (PC), and object motion states (OMS) constraints from part affordance graphs.
	}
	\resizebox{\columnwidth}{!}{%
		\begin{tabular}{l|cccccc}
			\hline
			        & VideoCLIP$\uparrow$ & Smoothness$\downarrow$ & Diversity$\uparrow$ & Non-collision$\uparrow$ & Contact$\uparrow$ \\
			\hline
			w/o PF  & \textbf{0.290}      & \textbf{0.004}         & \underline{0.81}    & \underline{0.99}        & \textbf{0.76}     \\
			w/o PC  & \underline{0.289}   & 0.011                  & 0.71                & \textbf{1.00}           & 0.26              \\
			w/o OMS & \textbf{0.290}      & \underline{0.006}      & 0.78                & \underline{0.99}        & \underline{0.73}  \\
			Ours    & \textbf{0.290}      & \textbf{0.004}         & \textbf{0.83}       & \underline{0.99}        & \textbf{0.76}     \\
			\hline
		\end{tabular}
	}
	\label{tab_sketchfab_ablations}
\end{table}

\begin{table}[t]
	\centering
	\caption{Evaluating different Large Language Models (LLMs) and Video Diffusion Models (VDMs) on Sketchfab.
         The performance of our implementation (based on DeepSeek and CogVideoX) remains stable when using a different LLM (Gemini) or VDM (HunyuanVideo).
         Results are averaged over multi-person single-object and single-person multi-object interaction generations. \vspace{-0.1cm}
		}
	\resizebox{\columnwidth}{!}{%
		\begin{tabular}{l|cccccc}
			\hline
			        & VideoCLIP$\uparrow$ & Smoothness$\downarrow$ & Diversity$\uparrow$ & Non-collision$\uparrow$ & Contact$\uparrow$ \\
			\hline
			LLM     &         0.291       &         0.004          &         0.73        &            0.99         &         0.68      \\
                VDM     &         0.289       &         0.002          &         0.81        &            0.99         &         0.76      \\
			Ours    &         0.290       &         0.004          &         0.83        &            0.99         &         0.76      \\
			\hline
		\end{tabular}
	}
	\vspace{-0.3cm}
	\label{tab_sketchfab_ablations_supp}
\end{table}
\begin{figure}[t]
    \centering
    \includegraphics[width=\columnwidth]{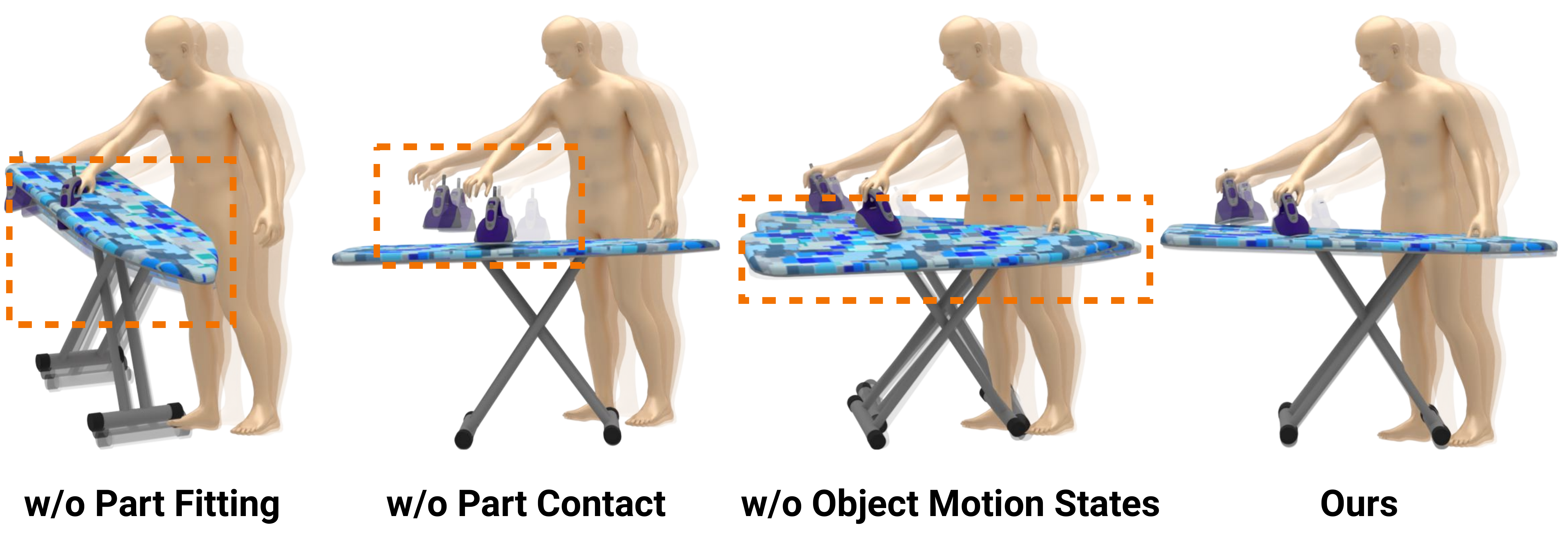}
    \caption{Visualization of ablation studies on part affordance graph constraints. Without part-level fitting, the ironing board orientation is incorrect (tilted up); without part-level contact, the hand is not in contact with the iron's handle; without object motion states, the ironing board does not remain stationary. Using all part affordance graph constraints produces the most realistic interactions.
    \vspace{-0.3cm}
    }
    \label{fig_sketchfab_ablation}
\end{figure}

\subsection{Ablation Studies}

\cref{tab_sketchfab_ablations} and \cref{fig_sketchfab_ablation} show the results of our ablation studies on the Sketchfab dataset.
We evaluate the effectiveness of our part affordance graph constraints:  
part-level fitting (\ie, $\LossObjectPartFit_{\text{3D}}, \LossObjectPartFit_{\text{2D}}$), part-level contact (\ie, $\LossContactContinuity$), and object motion states (\ie, $\GraphNodeRotational, \GraphNodeTranslational$ in $\LossSmoothness$).

\mypara{What is the impact of part-level fitting?}
Our part-level fitting (PF) during HOI optimization is essential for higher-level semantic plausibility not easily captured by standard quantitative metrics. Note that contact is measured at the whole body level, as we lack ground truth for part contacts. For instance, as shown in \cref{fig_sketchfab_ablation} (left), without part fitting, the ironing board has a wrongly tilted upwards orientation and significant motion, while using part fitting provides more meaningful semantic coherence.

\mypara{How do part contact constraints influence interaction quality?}
Without part-level contact constraints (w/o PC), high-level motions are plausible but miss important contacts. Our part contact constraints enable grasping of the iron handle with the person's hand in \cref{fig_sketchfab_ablation} (left middle).

\mypara{What is the effect of characterizing object motion states?}
Our characterization of object motion (OMS) in the PAG produces more semantically plausible object motion; for instance, this helps the ironing board remain stationary in \cref{fig_sketchfab_ablation}.

\mypara{How robust is our approach to the choice of foundation models?}
We test our approach with different Large Language Models (LLMs) and Video Diffusion Models (VDMs).
The default implementation uses DeepSeek as the LLM and CogVideoX as the VDM.
In \cref{tab_sketchfab_ablations_supp}, the first row reports the performance when swapping in Gemini as the LLM, while the second row reports the performance when swapping in Hunyuan-Video as the VDM, with results averaged over multi-person single-object and single-person multi-object interaction generations.
We observe that our approach achieves stable performance across the foundation models used.

\mypara{Limitations.}
Capturing detailed, nuanced motion (e.g., individual finger articulations) remains a challenge, lying beyond the granularity of our PAGs, which could be addressed through physics-based simulation.
Additionally, while our approach is robust to variations in the underlying foundation models, strong failures in these external components (e.g., consistently implausible first frames, video priors, or degenerate segmentation) can still degrade the final output quality.

\section{Conclusion}
\label{sec_conclusion}
We presented a new approach for zero-shot 4D human-object interaction synthesis that moves beyond whole-body interaction modeling by explicitly incorporating part-level affordances.
By introducing part affordance graphs, and grounding them to video motion generation and 4D HOI optimization, our method enables more realistic, diverse, and generalizable interactions across a wide range of objects and scenarios, including complex multi-object and multi-person interactions. 
We hope this step towards finer-grained understanding of interactions in a zero-shot fashion will open new possibilities in content creation as well as in applications such as robotics and embodied AI.

\mypara{Acknowledgements.}
This project is funded by the ERC Starting Grant SpatialSem (101076253), and the German Research Foundation (DFG) Grant ``Learning How to Interact with Scenes through Part-Based Understanding.''

\section*{Impact Statement}
Our method can benefit content creation, robotics, and embodied AI by enabling scalable synthesis of diverse human-object interactions without paired motion capture data. At the same time, the ability to generate plausible interactions carries a potential risk of misuse, as synthesized motions may misrepresent real human behaviors if presented without disclosure. We suggest clear labeling of synthetic content and caution when transferring generated interactions to safety-critical robotic systems.

\bibliography{reference}
\bibliographystyle{icml2026}

\newpage
\appendix
\onecolumn

In this appendix, we provide additional results in \cref{sec_supp_more_results} and more implementation details in \cref{sec_supp_implementation_details}.

\section{Additional Results}
\label{sec_supp_more_results}

\begin{figure}[h]
    \centering
    \includegraphics[width=0.9\columnwidth]{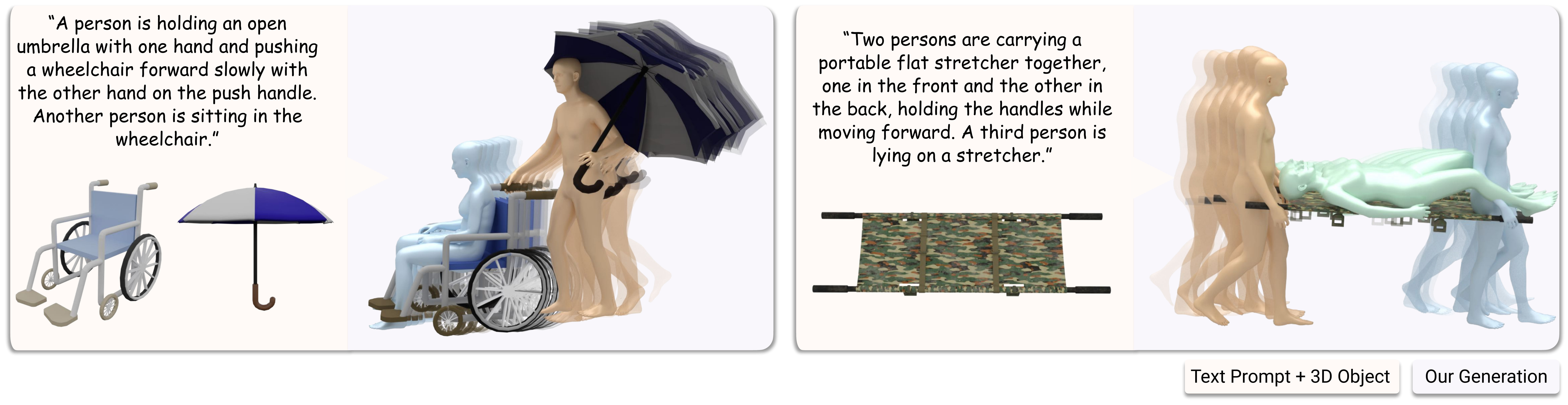}
    \caption{Our approach can generate multi-person multi-object interactions (\textbf{Left}) as well as interactions involving more than 2 people (\textbf{Right}).
    }
    \label{fig_sketchfab_more_multi_interaction}
\end{figure}

\mypara{Multi-person Multi-object Interaction Generation.}
\cref{fig_sketchfab_more_multi_interaction}-left demonstrates that our part affordance graph-based approach is flexible and can generate more complex multi-person multi-object interactions.
\cref{fig_sketchfab_more_multi_interaction}-right shows that our approach can generate interactions involving more than 2 people in a zero-shot fashion, going well beyond the single-person single-object interaction generation setting focused on in existing works \cite{peng2025hoidiff,li2024controllable}.

\begin{figure*}[ht]
    \centering
    \includegraphics[width=0.9\textwidth]{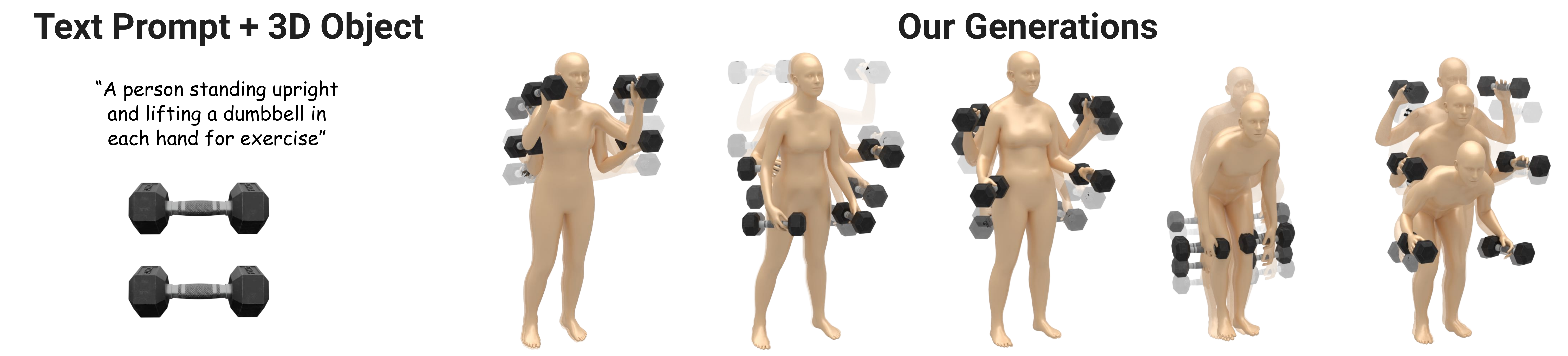}
    \vspace{-0.2cm}
    \caption{Our approach generates diverse 4D human-object interaction motions given the same text prompt and 3D objects as input.
    }
    \label{fig_sketchfab_diversity}
\end{figure*}

\mypara{Diversity Visualization.}
We visualize the generation diversity of our approach in \cref{fig_sketchfab_diversity}.
Given the same text prompt and 3D objects, our approach generates diverse 4D HOI interaction motions by varying the random noise used in video diffusion.

\mypara{Intermediate Result Visualization.}
\cref{fig_intermediate_results} presents intermediate results at different stages of our pipeline, including inferred part affordance graphs (\cref{subsec_method_part_affordance_graphs}), enhanced text prompts, 3D object part segmentation (\cref{subsec_method_multiview_object_part_segmentation}), interaction video generation, video object segmentation, depth estimation, and human motion recovery (\cref{subsec_method_hoi_video_generation}).

\mypara{Part-level Contact Metrics.}
To further evaluate the fine-grained physical realism of our 4D HOI generations, we compute part-level contact metrics.
Specifically, we sample points from human body parts and object part segmentations, and compute the minimum distance for each part-level contact.
Ground-truth part-level contacts are first identified by an LLM and then manually verified.
We evaluate two metrics: 
(1) \emph{Contact Accuracy}, the percentage of frames where part-level distances fall below a threshold $\tau$, and 
(2) \emph{Contact Distance}, the average part-level minimum distance across all frames.
As shown in \cref{tab_part_level_contact}, our approach achieves substantially higher part-level contact accuracy and much lower contact distance compared to both HOI-Diff and CHOIS, validating the superior physical realism of our generations.

\begin{table*}[h]
	\centering
	\caption{Part-level contact metrics on 4D human-object interaction generations. Our approach produces more accurate and tighter part-level contacts than the baselines HOI-Diff and CHOIS. \vspace{-0.1cm}
	}
	\resizebox{0.75\textwidth}{!}{%
		\begin{tabular}{l|ccc|c}
			\hline
			         & Contact Accuracy ($\tau$=1cm) $\uparrow$ & Contact Accuracy ($\tau$=3cm) $\uparrow$ & Contact Accuracy ($\tau$=5cm) $\uparrow$ & Contact Distance $\downarrow$ \\
			\hline
			HOI-Diff &             0.318                       &             0.434                       &             0.532                       &             0.144            \\
			CHOIS    &             0.274                       &             0.412                       &             0.486                       &             0.180            \\
			Ours     &     \textbf{0.887}                      &     \textbf{0.897}                      &     \textbf{0.906}                      &     \textbf{0.044}           \\
			\hline
		\end{tabular}
	}
	\label{tab_part_level_contact}
\end{table*}

\mypara{Evaluation on the BEHAVE Dataset.}
We further evaluate our approach on the BEHAVE dataset \cite{bhatnagar22behave}, which contains real-world object scans and HOI captures.
BEHAVE's test set has 18 objects. We sample 3 text prompts for each object and generate 5 interaction variations for each prompt.

We use the released models of HOI-Diff \cite{peng2025hoidiff} and CHOIS \cite{li2024controllable} for comparison.
Note that HOI-Diff was trained on BEHAVE, while CHOIS was trained on the FullBodyManipulation dataset \cite{li2023object}, which contains indoor object interaction captures similar to BEHAVE.
CHOIS is conditioned additionally on object waypoints, which are derived from our generation results.
The same set of metrics from \cref{sec_experiments}, including semantic alignment, temporal smoothness, motion diversity, and physical plausibility, are computed.

\begin{figure*}[h!]
    \centering
    \includegraphics[width=0.85\textwidth]{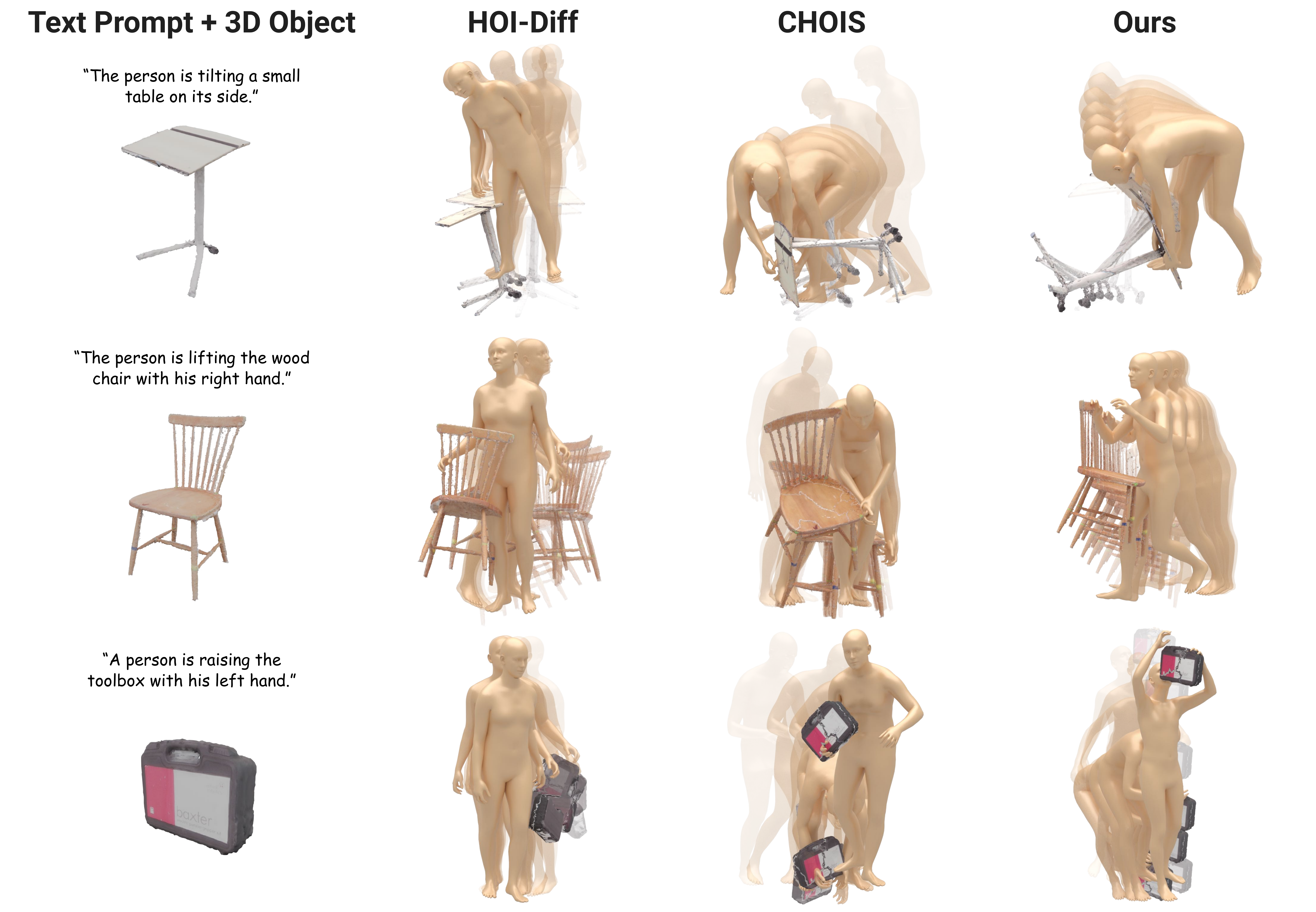}
    \vspace{-0.1cm}
    \caption{Qualitative comparisons of single-person single-object interaction generations on the BEHAVE dataset. Given real-world object scans and text prompts from BEHAVE, our 4D interaction generations align more closely with the text input than those of HOI-Diff \cite{peng2025hoidiff} and CHOIS \cite{li2024controllable}, which are specifically trained on captured data of real people interacting with such objects.
    }
    \label{fig_behave_qualitative_comparisons_single_interaction}
\end{figure*}

\cref{tab_behave_quantitative_comparisons} presents the quantitative comparisons, where our approach performs better than HOI-Diff and CHOIS in terms of semantic alignment, temporal smoothness, motion diversity, and physical contact.
\cref{fig_behave_qualitative_comparisons_single_interaction} shows the qualitative comparisons. Our approach synthesizes 4D interactions aligned more closely with text prompts than the generations of
HOI-Diff and CHOIS, which require captured interaction data for supervision.

\begin{figure*}[t]
    \centering
    \includegraphics[width=0.95\textwidth]{figures/intermediate_results_.pdf}
    \vspace{-0.1cm}
    \caption{Intermediate result visualization. Given 3D objects (e.g., a leather briefcase and a scooter) and a text prompt, we use an LLM to infer the part affordance graph (a).
    We also use the LLM to perform prompt enhancement (b) to capture the interaction details in (a) for video generation. We perform multi-view part segmentation (c) on the input 3D objects based on (a). Next, we generate an interaction video (d) guided by (b). We then detect, track, and segment objects and their parts in the video (e), estimate depth for each frame (f), and perform human motion recovery to estimate 4D human poses from the video.
    }
    \label{fig_intermediate_results}
\end{figure*}

\begin{table*}[t]
	\centering
	\caption{Comparing single-person single-object interaction generations on the BEHAVE dataset. Our approach achieves better performance than HOI-Diff \cite{peng2025hoidiff} and CHOIS \cite{li2024controllable} in semantic consistency, temporal smoothness, motion diversity, and physical contact metrics. \vspace{-0.1cm}
	}
	\resizebox{0.75\textwidth}{!}{%
		\begin{tabular}{l|c|cc|cc|cc}
			\hline
			         & Semantics            & \multicolumn{2}{c|}{Temporal Smoothness}                                         & \multicolumn{2}{c|}{Motion Diversity}                         & \multicolumn{2}{c}{Physical Plausibility}     \\
			         & VideoCLIP $\uparrow$ & Human $\downarrow$                       & Object $\downarrow$                   & Human $\uparrow$                          & Object $\uparrow$ & Non-collision $\uparrow$ & Contact $\uparrow$ \\
			\hline
			HOI-Diff &             0.200    &                                0.007     &                            0.015      &                                  0.34     &            0.54   &           \textbf{0.99}  &             0.72   \\
			CHOIS    &             0.214    &                                0.009     &                            0.008      &                                  0.48     &            0.47   &                   0.98   &             0.61   \\
			Ours     &     \textbf{0.220}   &                        \textbf{0.006}    &                    \textbf{0.004}     &                          \textbf{0.61}    &    \textbf{0.92}  &                   0.98   &     \textbf{0.78}  \\
			\hline
		\end{tabular}
	}
	\label{tab_behave_quantitative_comparisons}
\end{table*}

\begin{figure*}[h!]
    \centering
    \includegraphics[width=0.75\textwidth]{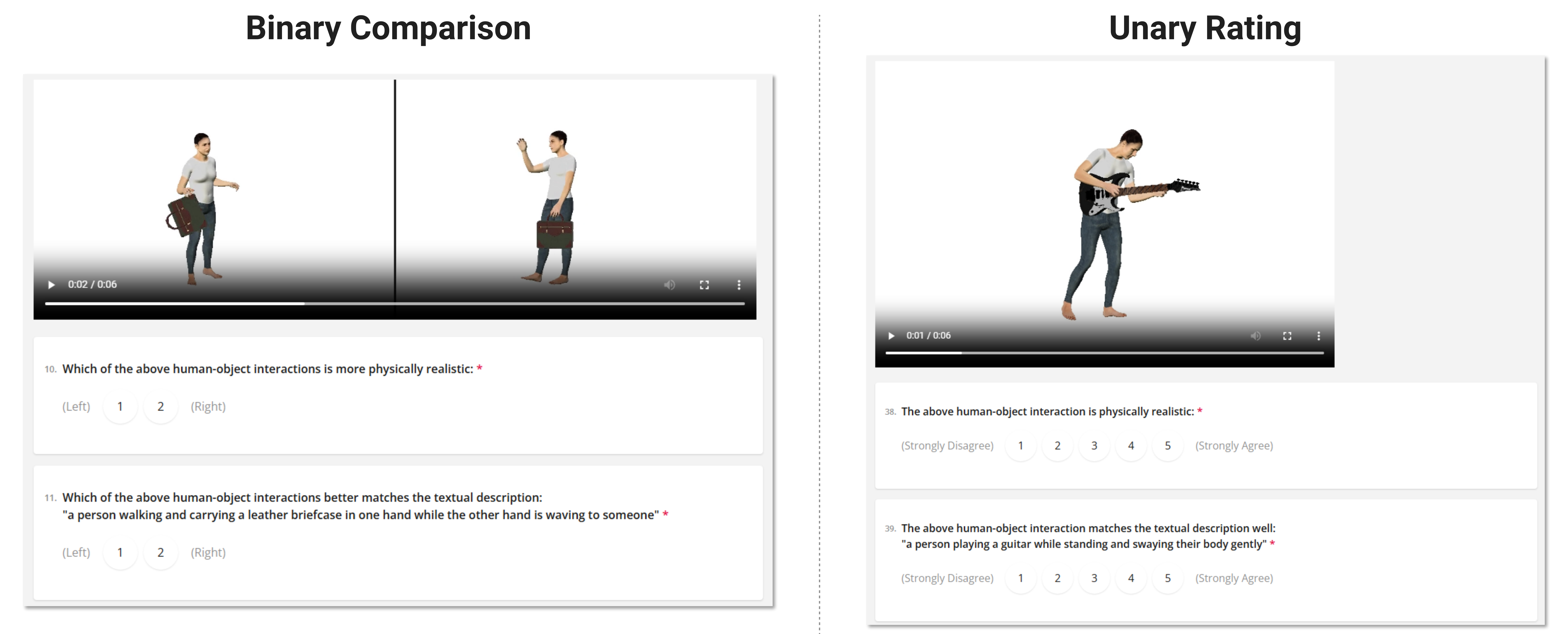}
    \vspace{-0.1cm}
    \caption{Screenshots of our perceptual study survey. Binary study (\textbf{Left}): participants are asked to select a 4D interaction generation with better realism and text alignment, respectively. Unary study (\textbf{Right}): rate generation realism and text alignment, respectively, on a scale from 1 to 5.}
    \label{fig_perceptual_study_screenshots}
\end{figure*}

\section{Additional Implementation Details}
\label{sec_supp_implementation_details}

\mypara{Point Map Alignment.}
To estimate point maps (or depth) for the generated video frames, we use MoGe~\cite{wang2024moge} due to its strong generalization to open-domain images and its more regularized 3D structure estimation (\cref{subsec_method_hoi_video_generation}).
However, MoGe is a single-image estimation method and suffers from inconsistencies across video frames.
Its point map estimation also does not align well with the 4D human motion estimated by GVHMR~\cite{shen2024gvhmr}. 
To address this, we perform a point map alignment step, leveraging the recovered 4D human motion as guidance.
We first detect and segment humans in the generated video frames, similar to Video Object Part Segmentation in (\cref{subsec_method_hoi_video_generation}).
We then optimize the scale, rotation, and translation of each point map frame so that the human point maps are aligned with the 4D human motion.
The optimization objective combines 3D and 2D fitting losses based on Chamfer distance, similar to $\LossObjectFit_{\text{3D}}$ and $\LossObjectFit_{\text{2D}}$ in \cref{subsec_method_hoi_optimization}.  
We perform 300 steps of gradient descent for this optimization.

\begin{table}[h!]
	\centering
	\caption{Runtime breakdown of our multi-stage pipeline. \vspace{-0.1cm}
	}
	\resizebox{\columnwidth}{!}{%
		\begin{tabular}{l|cccc}
			\hline
			            & LLM Reasoning & Object Segmentation & Video Generation \& Constraint Extraction & 4D HOI Optimization \\
			\hline
			Time (mins) &      0.6      &        5.7          &                   8.3                     &        6.0          \\
			\hline
		\end{tabular}
	}
	\label{tab_runtime}
\end{table}

\mypara{Runtime Analysis.}
\cref{tab_runtime} reports the runtime breakdown of our multi-stage pipeline.
Our approach remains unoptimized for efficiency, and could be further accelerated by employing 3D-native object segmentation models, faster video-generation models as they become available, and early stopping in the optimization.
Nevertheless, our approach is practical for an offline 4D synthesis system, particularly given its zero-shot methodology and the complexity of the 4D interaction output (multi-frame, multi-person/object).

\mypara{Perceptual Study.}
In our binary perceptual study, we have 14 generation comparisons, where each comparison consists of two questions: one for realism and one for text alignment.
In the unary study, participants are asked to rate 31 generations on realism and text alignment, respectively.
\cref{fig_perceptual_study_screenshots} shows the screenshots of our perceptual study survey.

\mypara{Prompting for Part Affordance Graph Inference.}
We provide the text prompt below for instructing an LLM~\cite{guo2025deepseek} to infer part affordance graphs (\cref{subsec_method_part_affordance_graphs}), while simultaneously enhancing short interaction prompts into longer, more detailed ones.

\begin{lstlisting}[
	basicstyle=\scriptsize\ttfamily, %
	emph={}, %
	emphstyle=\bfseries\color{black},
	frame=single,
  xleftmargin=0.3cm,
  xrightmargin=0.3cm,
	breaklines=true,
	backgroundcolor=\color{white},
	showstringspaces=false
]
You are a helpful assistant in analyzing human-object interactions. 

- Task: You will be given a list of objects and a short text description of human interactions with these objects. Your task is to analyze all the interaction relations among human body parts and object parts and output the results as a graph in the JSON format. 

- Input format: The input is provided in the JSON format as follows 
{
	"objects": [
		"object 1",
		"object 2"
	],
	"interaction": "a short interaction description"
} 

- Output format: Provide the output strictly in JSON format, without any additional explanation or commentary, structured as follows: 
{
	"object part nodes": [
		"object 1, object part 1",
		"object 1, object part 2"
	],
	"body part nodes": [
		"person 1, human body part 1",
		"person 1, human body part 2"
	],
	"interaction edges": [
		{
			"nodes": [
				"object a, object part b",
				"person c, human body part d"
			],
			"is_rel_static": <true or false indicating if the two nodes' movements remain relatively stationary during interaction>,
			"is_continuous": <true or false indicating if the two nodes remain in continuous physical contact during interaction>
		},
		{
			"nodes": [
				"object x, object part y",
				"person z, human body part w"
			],
			"is_rel_static": <true or false>,
			"is_continuous": <true or false>
		}
	],
	"interaction": "a long description in 150 words summarizing the output interaction graph to guide a realistic video generation",
	"object states": [
		{
			"name": "object 1",
			"is_translational": <true or false indicating if object 1 has translational motions during interaction>,
			"is_rotational": <true or false indicating if object 1 has rotational motions during interaction>,
			"description": "a short description in 20 words identifying object 1 during interaction"
		},
		{
			"name": "object 2",
			"is_translational": <true or false>,
			"is_rotational": <true or false>,
			"description": "a short description in 20 words identifying object 2 during interaction"
		}
	],
	"human states": [
		{
			"name": "person 1",
			"description": "a short description in 20 words identifying person 1 during interaction"
		}
	]
} 

- Rules for analysis: 
  (1) There are two types of nodes in the output interaction graph: "object part nodes" representing object parts and "body part nodes" representing human body parts. 
  (2) The "object part nodes" field represent a part-level segmentation of each input object. Segmentations should roughly cover the entire object without becoming excessively detailed. Use descriptive, specific part names rather than generic terms, for example, avoid "surface", "edge", "body", "base", "area", "cover", "support", "connector", "frame", and the like. Do not differentiate between left and right parts. Avoid numbering object parts. Example: For a "bike", use the following parts: "handlebar", "pedal", "seat", "frame tubes", "wheels". For a "skateboard", use the following parts: "longboard deck", "wheels". For a "cordless vacuum cleaner", use the following parts: "ergonomic hand grip", "wand", "floor roller". For a "ladder", use the following parts: "side rail tubes", "rungs". For a "boxing bag", use the following parts: "punching bag". 
  (3) The "body part nodes" field must be the following: "left hand", "right hand", "left arm", "right arm", "left shoulder", "right shoulder", "left leg", "right leg", "left foot", "right foot", "head", "hips". Distinguish between left/right human body parts. 
  (4) The "interaction edges" represent direct physical contact relationships between two end nodes. An edge connects an object part node to either a human body part node or another object part node. Do not connect part nodes within the same object. Example: when ironing on an ironing board, the soleplate part of an iron should be connected to the top flat panel part of the ironing board. Each edge has two attributes: "is_continuous" and "is_rel_static". The "is_continuous" attribute is true if the two end nodes are in continuous physical contact during the interaction process, otherwise false. Example: when holding a dumbbell, the hand is in continuous contact with the handle without any separation; when punching a boxing bag, the hands are not in continuous contact with the bag; when a person stepping up a ladder, the feet and hands are both not in continuous contact with the rungs. The "is_rel_static" attribute is true if the two end nodes' movements are relatively stationary to each other while being in continuous physical contact during the interaction process, otherwise false. Example: when riding a bike, hands are relatively stationary to the handlebar; when playing a guitar, the hand strumming strings is not relatively stationary to the main compartment of the guitar. 
  (5) Explicitly mentioned body parts in the input "interaction" field must be included. Example: For a description "a person is lifting a single dumbbell with one hand", include either "left hand" or "right hand" in the analysis. If no specific body part is mentioned, use the most common ergonomic interactions in the physical contact analysis. 
  (6) Focus on primary actions influencing object use or movement in the physical contact analysis. Example: For "a person walking and carrying a briefcase in one hand", the primary action for analysis is "carrying". 
  (7) Ensure the identified object parts belong to their respective objects in the node and edge outputs of the interaction graph. 
  (8) Ensure plausible distribution and avoid conflicts or duplication of human body parts during the interaction analysis. 
  (9) Exclude environmental elements, like floor, ground, or wall, from the physical contact analysis. 
  (10) The "interaction" field in the output JSON must concisely summarize the "interaction edges" of the graph to guide realistic video generation. Follow this structure: 
	(a) Begin with the interaction(s) as described in the input short "interaction" description. Clearly specify each participant's role if multiple people or objects are involved. All motions must occur at an extremely slow pace. 
	(b) Then describe the interaction motion details, focusing on physical contact between human body parts and object parts. If a human is specified to be non-static, make sure their body parts without physical contact show expressive movement. For example, when "skateboarding", the person's arms can swing to maintain balance, and the legs can bend slightly; when "cleaning with a cordless vacuum cleaner", the arm that is not holding the vacuum can swing naturally while walking; when "riding a scooter", one foot can remain static on the deck while the other swings to push off the ground and gain speed. Importantly, the human body parts without physical contact must also move in slow motion. 
	(c) Next, describe the appearance of people, objects, and environments. For people, you must strictly include the following four aspects: their hair styles, facial expressions, clothes, and shoes. For example, "short black hair", "neutral facial expression", "wearing a gray shirt, blue jeans, and white sneakers". For objects, describe general type and appearance without overly specific details. The environment is always a clean, spacious indoor area with white walls and a wooden floor. Ensure the environment supports the action without adding unnecessary complexity. 
	(d) The "interaction" summarization must not exceed 150 words. 
  (11) The "object states" in the output JSON have four attributes, "name", "is_translational", "is_rotational", and "description", for each object. The "is_translational" attribute is true if the corresponding object has global translational motions during interaction, otherwise false. The "is_rotational" attribute is true if the corresponding object has global rotational motions during interaction, otherwise false. Both "is_translational" and "is_rotational" attributes must consider only the object's overall motion, not motions of individual parts, for example, a bike being ridden should be considered as moving translationally as a whole, while ignoring the rotation of its pedals. The object "description" attribute should clearly identify the object by briefly stating its type, appearance, and its interactions with human bodies, using no more than 20 words. The object "description" should be based on relevant "interaction edges" and the long "interaction" fields in the output. In the object "description", avoid using numerical or ordinal references. 
  (12) The "human states" in the output JSON have two attributes, "name" and "description", for each person. The human "description" attribute should clearly identify the person by briefly stating their appearance and interactions with object parts in 20 words. The human "description" should be based on relevant "interaction edges" and the long "interaction" fields in the output. Avoid using numerical or ordinal references in the "description" attribute. 

- Examples:
  (1) If the input is 
{
	"objects": [
		"umbrella",
		"suitcase"
	],
	"interaction": "a person is dragging a suitcase with one hand and holding an open umbrella with the other hand while walking"
} 
  then the output is 
{
	"object part nodes": [
		"umbrella, canopy",
		"umbrella, shaft",
		"suitcase, main compartment",
		"suitcase, handle",
		"suitcase, wheels"
	],
	"body part nodes": [
		"person 1, left hand",
		"person 1, right hand",
		"person 1, left arm",
		"person 1, right arm",
		"person 1, left shoulder",
		"person 1, right shoulder",
		"person 1, left leg",
		"person 1, right leg",
		"person 1, left foot",
		"person 1, right foot",
		"person 1, head",
		"person 1, hips"
	],
	"interaction edges": [
		{
			"nodes": [
				"umbrella, shaft",
				"person 1, left hand"
			],
			"is_rel_static": true,
			"is_continuous": true
		},
		{
			"nodes": [
				"suitcase, handle",
				"person 1, right hand"
			],
			"is_rel_static": true,
			"is_continuous": true
		}
	],
	"interaction": "A person is dragging a suitcase's handle with the right hand and holding a open umbrella's shaft with the left hand while walking at a slow pace. The suitcase rolls smoothly behind them as they move, and the open umbrella is held steadily above. The person has black short hair and a neutral facial expression. They wear a gray shirt, blue jeans, and white sneakers. The scene takes place in a clean, spacious indoor area with white walls and a wooden floor.",
	"object states": [
		{
			"name": "umbrella",
			"is_translational": true,
			"is_rotational": false,
			"description": "the open umbrella being held"
		},
		{
			"name": "suitcase",
			"is_translational": true,
			"is_rotational": false,
			"description": "the suitcase being dragged"
		}
	],
	"human states": [
		{
			"name": "person 1",
			"description": "the person with black short hair who is wearing gray shirt and blue jeans and holding/dragging the objects"
		}
	]
} 

  (2) If the input is 
{
	"objects": [
		"bike"
	],
	"interaction": "a person is riding a bike"
} 
  then the output is 
{
	"object part nodes": [
		"bike, handlebar",
		"bike, pedal",
		"bike, seat",
		"bike, frame tubes",
		"bike, wheels"
	],
	"body part nodes": [
		"person 1, left hand",
		"person 1, right hand",
		"person 1, left arm",
		"person 1, right arm",
		"person 1, left shoulder",
		"person 1, right shoulder",
		"person 1, left leg",
		"person 1, right leg",
		"person 1, left foot",
		"person 1, right foot",
		"person 1, head",
		"person 1, hips"
	],
	"interaction edges": [
		{
			"nodes": [
				"bike, handlebar",
				"person 1, left hand"
			],
			"is_rel_static": true,
			"is_continuous": true
		},
		{
			"nodes": [
				"bike, handlebar",
				"person 1, right hand"
			],
			"is_rel_static": true,
			"is_continuous": true
		},
		{
			"nodes": [
				"bike, pedal",
				"person 1, left foot"
			],
			"is_rel_static": true,
			"is_continuous": true
		},
		{
			"nodes": [
				"bike, pedal",
				"person 1, right foot"
			],
			"is_rel_static": true,
			"is_continuous": true
		},
		{
			"nodes": [
				"bike, seat",
				"person 1, hips"
			],
			"is_rel_static": true,
			"is_continuous": true
		}
	],
	"interaction": "A person is riding a bike at a slow, steady pace in a clean, spacious indoor area with white walls and a wooden floor. Their hands grip the handlebars firmly and feet remain securely on the pedals. The bike has a simple, modern design with a black frame and straight handlebars. The rider has short brown hair and a neutral facial expression. They wear a blue shirt, black shorts, and white sneakers.",
	"object states": [
		{
			"name": "bike",
			"is_translational": true,
			"is_rotational": false,
			"description": "the bike having a black frame and being ridden"
		}
	],
	"human states": [
		{
			"name": "person 1",
			"description": "the person who is wearing blue shirt and black shorts and riding"
		}
	]
} 

  (3) If the input is 
{
	"objects": [
		"guitar"
	],
	"interaction": "a person is playing a guitar while standing"
} 
  then the output is 
{
	"object part nodes": [
		"guitar, neck",
		"guitar, main compartment"
	],
	"body part nodes": [
		"person 1, left hand",
		"person 1, right hand",
		"person 1, left arm",
		"person 1, right arm",
		"person 1, left shoulder",
		"person 1, right shoulder",
		"person 1, left leg",
		"person 1, right leg",
		"person 1, left foot",
		"person 1, right foot",
		"person 1, head",
		"person 1, hips"
	],
	"interaction edges": [
		{
			"nodes": [
				"guitar, neck",
				"person 1, left hand"
			],
			"is_rel_static": false,
			"is_continuous": true
		},
		{
			"nodes": [
				"guitar, main compartment",
				"person 1, right hand"
			],
			"is_rel_static": false,
			"is_continuous": true
		}
	],
	"interaction": "A person is playing a guitar while standing in a clean, spacious indoor area with white walls and a wooden floor. Their left hand is holding the guitar's fretboard, and their right hand is strumming the strings slowly. The guitar is a classic acoustic model with a polished wood finish. The person has short brown hair and a happy faical expression. They wear a black shirt, blue jeans, and black boots, gently swaying their body to the rhythm.",
	"object states": [
		{
			"name": "guitar",
			"is_translational": true,
			"is_rotational": false,
			"description": "the wooden guitar being played"
		}
	],
	"human states": [
		{
			"name": "person 1",
			"description": "the person with short brown hair who is wearing blue jeans and playing the guitar"
		}
	]
} 
\end{lstlisting}

\mypara{Prompting for First-Frame Selection.}
The following text prompt is used to instruct a VLM (GPT-4.1) to select the best first frame from a candidate set (\cref{subsec_method_implementation}) for video diffusion.

\begin{lstlisting}[
	basicstyle=\scriptsize\ttfamily, %
	emph={}, %
	emphstyle=\bfseries\color{black},
	frame=single,
  xleftmargin=0.3cm,
  xrightmargin=0.3cm,
	breaklines=true,
	backgroundcolor=\color{white},
	showstringspaces=false
]
You are a helpful assistant in image understanding and comparison. 
- Task: You will receive one image file that actually contains two separate images shown side-by-side (left and right), along with a short text describing human-object interactions. Look closely at both images and read the text description. Use the "Analysis Rules" below to decide which single image ("left" or "right") is a better match for both the rules and the text description. 
- Input format: 
	(1) One image file that includes two images placed next to each other horizontally, like this: [left image | right image]. 
	(2) One short text that describes the human-object interactions that should be happening in the images. 
- Output format: You must output only one word: either "left" or "right". Do not add any other words, explanations, or comments. 
- Analysis Rules: 
	(1) Full Human Figures: Prefer the image where people are shown completely, from their heads down to their feet, inside the image area, and where the front faces of the main people involved in the interaction are clearly visible. 
	(2) Correct Anatomy: Prefer the image where humans have normal-looking body parts and proportions. Avoid images showing people with distorted, disfigured, or anatomically incorrect limbs or bodies. 
	(3) Matching Text Description: Prefer the image where the human-object interactions match the provided short text description. 
	(4) Plausible Interactions: Prefer the image where interactions between people and objects look natural, physically plausible. Avoid interactions that involve problematic body parts, like strangely bent or extra limbs. Avoid images with unrealistic physics, like people or objects floating in the air. 
	(5) Camera View: Prefer wide-shot images taken from a shoulder-height, three-quarter side view that clearly shows both the pose and the interaction. If that's not available, prefer side views over straight-on front views. Avoid images taken from high-up, low-down, or close-up views that crop or obscure full human figures. Also avoid images where people or objects are too close to walls or background objects. 
	(6) Sharp Details: Prefer images with clear, sharp details, and avoid images with motion blur around human body parts. 
	(7) Realistic Style: Prefer photographic or realistic images over cartoons, drawings, illustrations, or images with very artistic styles. 
	(8) Do not consider the mood, feeling, or atmosphere of the image in your comparison. 
\end{lstlisting}

\mypara{LLM Usage Disclosure.}
LLMs (ChatGPT and Gemini) were used for correcting grammatical errors and typos and finding synonyms in paper writing.

\mypara{Data Acknowledgements.}
We collected 24 object models from Sketchfab.com for our experiments.

The following models are licensed under \href{http://creativecommons.org/licenses/by/4.0/}{Creative Commons Attribution}:
\begin{itemize}[leftmargin=1.5em, labelwidth=1.5em, labelsep=0.5em, topsep=-0.5em, itemsep=0em, parsep=0em]
    \item \href{https://skfb.ly/oExAn}{Army Stretcher} by 4mecharmi,
    \item \href{https://skfb.ly/oyZ6w}{Bicycle Game Asset} by RayznGames,
    \item \href{https://skfb.ly/oJYwt}{Briefcase} by Artistic7,
    \item \href{https://skfb.ly/oupLE}{Barbell} by Bluups,
    \item \href{https://skfb.ly/oLvVs}{Boxing Bag} by Francisco Alvarez Mendez,
    \item \href{https://skfb.ly/oCIA9}{Cruising Canoe} by gogiart,
    \item \href{https://skfb.ly/ouOvG}{Captain America's Shield} by A.I.R,
    \item \href{https://skfb.ly/onqXo}{Clothes Basket} by eeelabvisual,
    \item \href{https://skfb.ly/6RBZw}{Electric Scooter} by Gest.lt,
    \item \href{https://skfb.ly/6WXOz}{Harp} by neutralize,
    \item \href{https://skfb.ly/6RJsB}{ibanez jem guitar} by abazibiz,
    \item \href{https://skfb.ly/oKr7o}{Ironing Board with Iron} by GeniusPilot2016,
    \item \href{https://skfb.ly/oJT9M}{Lawn mower LP} by I\_am\_ball,
    \item \href{https://skfb.ly/oOpIt}{microphone} by ssmilerok,
    \item \href{https://skfb.ly/oUEyX}{Rocking Chair} by Dimension Dazzle,
    \item \href{https://skfb.ly/6VQq7}{Suitcase} by ekin,
    \item \href{https://skfb.ly/l4j3hd0a}{skateboard} by Chaitanya Krishnan,
    \item \href{https://skfb.ly/6YpNI}{Umbrella} by Diccbudd,
    \item \href{https://skfb.ly/oKMVI}{Vacuumed Cleaner} by Panda,
    \item \href{https://skfb.ly/puUGY}{Wheelchair} by Dodoyaco.
\end{itemize}

The following models are licensed under \href{http://creativecommons.org/licenses/by-sa/4.0/}{Creative Commons Attribution-ShareAlike}:
\begin{itemize}[leftmargin=1.5em, labelwidth=1.5em, labelsep=0.5em, topsep=-0.5em, itemsep=0em, parsep=0em]
    \item \href{https://skfb.ly/ornAq}{Hex Dumbbell 10kg} by Salim Rached,
    \item \href{https://skfb.ly/6XKs9}{Wheelbarrow} by Hene.
\end{itemize}

The following models are licensed under \href{https://sketchfab.com/licenses}{Free Standard License}:
\begin{itemize}[leftmargin=1.5em, labelwidth=1.5em, labelsep=0.5em, topsep=-0.5em, itemsep=0em, parsep=0em]
    \item \href{https://skfb.ly/o6pIK}{Modern Iron} by assetfactory,
    \item \href{https://skfb.ly/o9QwE}{Stool 02} by Nichgon.
\end{itemize}

\end{document}